\documentstyle[aps]{revtex}
\begin{document}
\draft
\title{Canonical Transformations and the Hamilton-Jacobi Theory 
 		      in Quantum Mechanics}
\author{Jung-Hoon Kim\footnote{e-mail: jhkim@laputa.kaist.ac.kr}
         and Hai-Woong Lee}
\address{
         Department of Physics,
         Korea Advanced Institute of Science and Technology,
         Taejon, 305-701, Korea }
\maketitle
\begin{abstract}
Canonical transformations using the idea of quantum generating functions
are applied to construct a quantum Hamilton-Jacobi theory, based on the 
analogy with the classical case. An operator and a c-number forms of the 
time-dependent quantum Hamilton-Jacobi equation are derived and used to 
find dynamical solutions of quantum problems. The phase-space picture of 
quantum mechanics is discussed in connection with the present theory.
\end{abstract}

\section{Introduction}

Various mechanical problems can be elegantly approached by the Hamiltonian
formalism, which not only found well-established ground in classical 
theories\cite{one}, but also provided much physical insight in the early 
development of quantum theories\cite{two,three}. It is curious though that 
the concept of canonical transformations, which plays a fundamental role in 
the Hamiltonian formulation of classical mechanics, has not attracted as
much attention in the corresponding formulation of quantum mechanics. A 
relatively small quantity of literature is available as of now on 
this subject [4--10]. 
The main reason for this is probably that canonical variables in quantum
mechanics are not c-numbers but noncommuting operators, manipulation of which
is considerably involved. In spite of this difficulty, the great success of 
canonical transformations in classical mechanics makes it desirable to 
investigate the possibility of application of the concept of canonical 
transformations in quantum mechanics at least to the extent allowed in view 
of the analogy with the classical case.

The usefulness of the classical canonical transformations is most visible in 
the Hamilton-Jacobi theory where one seeks a generating function that makes
the transformed Hamiltonian become identically zero\cite{one}. A quantum 
analog of the Hamilton-Jacobi theory has previously been considered by 
Leacock and Padgett\cite{eight} with particular emphasis on the quantum 
Hamilton's characteristic function and applied to the definition of the
quantum action variable and the determination of the bound-state energy 
levels\cite{one2}. However, the {\em dynamical} aspect of the quantum
Hamilton-Jacobi theory appears to remain untouched. In the present study, 
we concentrate on this aspect of the problem, and derive the time-dependent
quantum Hamilton-Jacobi equation following closely the procedure that lead 
to the classical Hamilton-Jacobi equation. 

The analogy between the classical and quantum Hamilton-Jacobi theories can
be best exploited by employing the idea of the quantum generating function 
that was first introduced by Jordan\cite{four} and Dirac\cite{five}, and 
recently reconsidered by Lee and l'Yi\cite{ten}. The ``well-ordered'' 
operator counterpart of the quantum generating function is used in 
constructing our quantum Hamilton-Jacobi equation, which resembles in form
the classical Hamilton-Jacobi equation. By means of well-ordering, a unique 
operator is associated with a given c-number function, thereby the ambiguity 
in the ordering problem is removed. We identify the quantum generating 
function accompanying the quantum Hamilton-Jacobi theory as the quantum 
Hamilton's principal function, and apply this theory to find the dynamical 
solutions of quantum problems.

The prevailing conventional belief that physical observables should be 
Hermitian operators invokes in our discussion the unitary transformation 
that transforms one Hermitian operator to another. This along with the 
fact that the unitary transformation preserves the fundamental quantum 
condition for the new canonical variables $[\hat{Q},\hat{P}]=i\hbar$ if 
the old canonical variables satisfy $[\hat{q},\hat{p}]=i\hbar$ provides a 
good reason why we call the unitary transformation the quantum canonical 
transformation. This definition of the quantum canonical transformation is
analogous to the classical statement that the classical canonical 
transformation keeps the Poisson brackets invariant, i.e., 
$[Q,P]_{PB}=[q,p]_{PB}=1$. In our current discussion of the quantum 
canonical transformation we will consider exclusively the case of the 
unitary transformation. 

The paper is organized as follows. In Sec.\ II the quantum canonical 
transformation using the idea of the quantum generating function is briefly 
reviewed, and the transformation relation between the new Hamiltonian and 
the old Hamiltonian expressed in terms of the quantum generating function is
derived. From this relation, and by analogy with the classical case, we 
arrive at the quantum Hamilton-Jacobi equation in Sec.\ III. It will be found
that the unitary transformation of the special type 
$\hat{U}(t)=\hat{T}(t)\hat{A}$ where $\hat{T}(t)$ is the time-evolution 
operator and $\hat{A}$ is an arbitrary time-independent unitary operator 
satisfies the quantum Hamilton-Jacobi equation. 
Sec.\ IV is devoted to the discussion of 
the quantum phase-space distribution function under canonical transformations. 
The differences between our approach and that of Ref.\cite{one1} are described.
Boundary conditions and simple applications of the theory are given 
in Sec.\ V, where to perceive the main idea easily most of the discussion 
is developed with the simple case $\hat{A}=\hat{I}$, the unit operator, while
keeping in mind that the present formalism is not restricted to this case.
Finally, Sec.\ VI presents concluding remarks.

\section{Quantum Canonical Transformations}

Let us begin our discussion by reviewing the theory of the quantum canonical
transformations\cite{five,ten}. A quantum generating function that is 
analogous to a classical generating function is defined in terms of the 
matrix elements of a unitary operator as follows\cite{five},
\begin{equation}
e^{iF_1(q_1,Q_2,t)/\hbar}\equiv \langle q_1|Q_2\rangle _t
=\langle q_1|\hat{U}(t)|q_2\rangle ,
\end{equation}
where the unitary operator $\hat{U}(t)$ transforms an eigenvector of 
$\hat{q}$ into an eigenvector of $\hat{Q}=\hat{U}\hat{q}\hat{U}^\dagger$, 
i.e., $|Q_1\rangle _t=\hat{U}(t)|q_1\rangle$ (and $|P_1 
\rangle _t=\hat{U}(t)|p_1\rangle$).\footnote{
An eigenvalue $X_1$ and an eigenvector $|X_1\rangle$ of an
operator $\hat{X}$ are defined by the equation, $\hat{X}|X_1\rangle 
=X_1|X_1\rangle$ ($X=q$, $p$, $Q$, and $P$).
Different subindices are used to distinguish different eigenvalues or 
eigenvectors, e.g., $X_2$, $|X_2\rangle$, etc.; The subscript $t$ on a
ket $|\rangle _t$ (bra ${_t\langle}|$) expresses time dependence of the
ket $|\rangle _t$ (bra ${_t\langle}|$).} 
Different types of the quantum 
generating function can be defined similarly\cite{ten}, i.e.,
$e^{iF_2(q_1,P_2,t)/\hbar}\equiv \langle q_1|P_2\rangle _t
=\langle q_1|\hat{U}(t)|p_2\rangle$, 
$e^{iF_3(p_1,Q_2,t)/\hbar}\equiv \langle p_1|Q_2\rangle _t
=\langle p_1|\hat{U}(t)|q_2\rangle$, and
$e^{iF_4(p_1,P_2,t)/\hbar}\equiv \langle p_1|P_2\rangle _t
=\langle p_1|\hat{U}(t)|p_2\rangle$.

The quantum canonical transformation, or the unitary transformation, 
corresponds to a change of representation or equivalently to a rotation 
of axes in the Hilbert space.  The unitary transformation guarantees that 
the fundamental quantum condition $[\hat{Q},\hat{P}]
 =[\hat{q},\hat{p}]=i\hbar$ holds, the new canonical variables 
$(\hat{Q},\hat{P})$ are Hermitian operators, and the eigenvectors of 
$\hat{Q}$ or $\hat{P}$ form a complete basis. One should keep in mind 
that the eigenvalue $Q_1$ has the same numerical value as the eigenvalue 
$q_1$ because the unitary transformation preserves the eigenvalue spectrum
of an operator\cite{three}. In cases where it is convenient, one is free to 
interchange $q_1$ $ (p_1)$ with $Q_1$ $(P_1)$. 

Transformation relations between $(\hat{q},\hat{p})$ and $(\hat{Q},\hat{P})$ 
can be expressed in terms of the ``well-ordered'' generating operator 
$\bar{F}_1(\hat{q},\hat{Q},t)$\cite{five} that is an operator counterpart 
of the quantum generating function $F_1(q_1,Q_2,t)$ as follows:\footnote{
A well-ordered operator $\bar{G}(\hat{X},\hat{Y})$ is 
developed from a c-number function $G(X_1,Y_2)$ such that 
$\langle X_1|\bar{G}(\hat{X},\hat{Y})|Y_2
\rangle = G(X_1,Y_2)\langle X_1|Y_2\rangle $\cite{five}. For example,
if $G(X_1,Y_2)=X_1Y_2+Y_2^2X_1^3$, then $\bar{G}(\hat{X},\hat{Y})=
\hat{X}\hat{Y}+\hat{X}^3\hat{Y}^2$.}
\begin{equation}
\hat{p}=\frac{\partial \bar{F}_1(\hat{q},\hat{Q},t)}{\partial \hat{q}}, 
\hspace{1cm} 
\hat{P}=-\frac{\partial \bar{F}_1(\hat{q},\hat{Q},t)}{\partial \hat{Q}}.
\end{equation}
Similar expressions for other types of the generating operators can be 
immediately inferred by analogy with the classical relations. For a later 
reference, we present the relations for $\bar{F}_2(\hat{q},\hat{P},t)$ below,
\begin{equation}
\hat{p}=\frac{\partial \bar{F}_2(\hat{q},\hat{P},t)}{\partial \hat{q}}, 
\hspace{1cm} 
\hat{Q}=\frac{\partial \bar{F}_2(\hat{q},\hat{P},t)}{\partial \hat{P}}.
\end{equation}

It is interesting to note that, whereas the four types of the generating 
functions in classical mechanics are related with each other through the 
Legendre transformations\cite{one}, the relations between the quantum
generating functions of different types can be expressed by means of the 
Fourier transformations. For example, the transition from 
$F_1(q_1,Q_2,t)$ to $F_2(q_1,P_2,t)$ can be accomplished by 
\begin{eqnarray}
e^{iF_2(q_1,P_2,t)/\hbar}&=&\int dQ_2 \langle q_1|Q_2\rangle _t\hspace{0.7mm} 
{_t\langle} Q_2|P_2\rangle _t, \nonumber \\
&=&\frac{1}{\sqrt{2\pi \hbar}} \int dQ_2 e^{iF_1(q_1,Q_2,t)/\hbar}
e^{iP_2Q_2/\hbar}.
\end{eqnarray}

The usefulness of the concept of the quantum generating function can be 
revealed, for example, by considering the unitary transformation
$\hat{U} =e^{ig(\hat{q})/\hbar}$ where $g$ is an arbitrary real function.
From the definition of the quantum generating function, we have
\begin{eqnarray}
e^{iF_2(q_1,P_2)/\hbar}&=&\langle q_1|e^{ig(\hat{q})/\hbar}|p_2 \rangle ,
\nonumber \\
&=& \frac{1}{\sqrt{2\pi \hbar}}e^{\frac{i}{\hbar}[g(q_1)+q_1P_2]}.
\end{eqnarray}
The well-ordered generating operator is then given by
\begin{equation}
\bar{F}_2(\hat{q},\hat{P})=g(\hat{q})+\hat{q}\hat{P}+i\frac{\hbar}{2} 
\ln 2\pi \hbar ,
\end{equation}
and Eq.\ (3) yields the transformation relations
\begin{eqnarray}
\hat{Q}&=& \hat{q}, \\
\hat{P}&=&\hat{p}-\frac{\partial g(\hat{q})}{\partial \hat{q}}.
\end{eqnarray}
This shows that, in some cases, an introduction of the quantum generating 
function can provide an effective method of finding the transformation
relations between ($\hat{q},\hat{p}$) and ($\hat{Q},\hat{P}$) without 
recourse to the equations $\hat{Q}=\hat{U}\hat{q}\hat{U}^\dagger$ and 
$\hat{P}=\hat{U}\hat{p}\hat{U}^\dagger$.

Now we consider the dynamical equations governing the time-evolution of 
quantum systems. The time-dependent Schr\"{o}dinger equation for the system 
with the Hamiltonian $H(\hat{q},\hat{p},t)$ is given in terms of a 
time-dependent ket $|\psi \rangle _t$ by
\begin{equation}
i\hbar \frac{\partial}{\partial t}|\psi \rangle _t
=H(\hat{q},\hat{p},t)|\psi \rangle _t.
\end{equation}
In $Q$-representation the time-dependent Schr\"{o}dinger equation takes
the form
\begin{equation}
i\hbar \frac{\partial}{\partial t}\psi ^Q(Q_1,t)=K\left( Q_1,
-i\hbar \frac{\partial}{\partial Q_1},t\right) \psi ^Q(Q_1,t),
\end{equation}
where $\psi ^Q(Q_1,t)={_t\langle}Q_1|\psi \rangle _t$, and
\begin{equation}
K(\hat{Q},\hat{P},t)=H(\hat{q},\hat{p},t)+i\hbar \hat{U}
\frac{\partial \hat{U}^\dagger}{\partial t}.
\end{equation}
The second term on the right hand side of Eq.\ (11) arises from the fact 
that we allow the time dependence of the unitary operator $\hat{U}(t)$, 
which indicates that, even though we adopt here the Schr\"{o}dinger picture 
where the time dependence associated with the dynamical evolution of a 
system is attributed solely to the ket $|\psi \rangle _t$, $\hat{Q}$ and 
$|Q_1\rangle _t$ may depend on time also.  In terms of the generating 
operator $\bar{F}_1(\hat{q},\hat{Q},t)$, Eq.\ (11) can be written as
\begin{equation}
K(\hat{Q},\hat{P},t)=H(\hat{q},\hat{p},t)+\frac{\partial \bar{F} _1
(\hat{q},\hat{Q},t)}{\partial t}.
\end{equation}
The equivalence of Eqs.\ (11) and (12) can be proved as shown in Appendix A.
It is important to note that $K(\hat{Q},\hat{P},t)$ plays the role of the 
transformed Hamiltonian governing the time-evolution of the system in 
$Q$-representation. The analogy with the classical theory is remarkable.

\section{Quantum Hamilton-Jacobi Theory}

We are now ready to proceed to formulate the quantum Hamilton-Jacobi theory.
One can immediately notice that, if $K(\hat{Q},\hat{P},t)$ of Eq.\ (12) 
vanishes, the time-dependent Schr\"{o}dinger equation in $Q$-representation 
yields a simple solution, $\psi ^Q=$\ const. This observation along with 
Eq.\ (2) naturally leads us to the following quantum Hamilton-Jabobi equation,
\begin{equation}
H\left( \hat{q},\frac{\partial \bar{S}_1(\hat{q},\hat{Q},t)}{\partial 
\hat{q}},t\right) +\frac{\partial \bar{S}_1(\hat{q},\hat{Q},t)}{\partial t} =0,
\end{equation}
where, following the classical notational convention, we denote the 
generating operator that is analogous to the classical Hamilton's principal 
function by $\bar{S}_1(\hat{q},\hat{Q},t)$. Eq.\ (13) bears a close formal 
resemblance to the classical Hamilton-Jacobi equation. It, however, differs 
from the classical equation in that it is an operator partial differential 
equation. The procedure of solving dynamical problems is completed if we 
express the wave function in the original $q$-representation as
\begin{eqnarray}
\psi ^q(q_1,t)&=&\int \langle q_1|Q_2 \rangle _t\hspace{0.7mm} 
{_t\langle}Q_2| \psi \rangle _t dQ_2, \nonumber \\ 
&=&\int e^{iS_1(q_1,Q_2,t)/\hbar}\psi ^Q(Q_2)dQ_2,
\end{eqnarray}
where $S_1(q_1,Q_2,t)$ is the c-number counterpart of 
$\bar{S}_1(\hat{q},\hat{Q},t)$, and is obtained by replacing the well-ordered 
$\hat{q}$ and $\hat{Q}$ in $\bar{S}_1$, respectively, with $q_1$ and $Q_2$. 
In Eq.\ (14), $t$ is dropped from $\psi ^Q$, since 
${_t\langle}Q_2| \psi \rangle _t =$ const. 
As is the case for the classical Hamilton-Jacobi equation, the mission of
solving dynamical problems is assigned to the quantum Hamilton-Jacobi
equation.

Even though we arrive at the correct form of the quantum Hamilton-Jacobi 
equation, it seems at first sight quite difficult to attain solutions of it 
due to its unfamiliar appearance as an operator partial differential equation.
Thus it seems desirable to search a corresponding c-number form of the 
quantum Hamilton-Jacobi equation. For this task, we note that, if the unitary
operator $\hat{U}(t)$ is assumed to be separable into 
$\hat{U}(t)=\hat{T}(t)\hat{A}$, where $\hat{T}(t)$ is the time-evolution 
operator 
and $\hat{A}$ is an arbitrary time-independent unitary operator, then 
$\psi ^Q(Q_1,t) ={_t\langle}Q_1|\psi \rangle _t=\langle q_1|\hat{A}^\dagger 
\hat{T}^\dagger (t)\hat{T}(t)|\psi (t=0) \rangle =\langle q_1|\hat{A}^\dagger 
|\psi (t=0)\rangle =$\ const. This means that the left hand side of Eq.\ (10)
becomes zero, i.e., the canonical transformation mediated by a separable 
unitary operator is exactly the one that we seek. Assuming 
$\hat{U}(t)=\hat{T}(t)\hat{A}$, we rewrite Eq.\ (1) as
\begin{equation}
e^{iS_1(q_1,Q_2,t)/\hbar} =\langle q_1|\hat{T}(t)\hat{A}|q_2\rangle . 
\end{equation}
Differentiating this equation with respect to time, we obtain
\begin{eqnarray}
\frac{i}{\hbar}\frac{\partial S_1}{\partial t}e^{iS_1/\hbar} &=&
\langle q_1 |\frac{\partial \hat{T}}{\partial t}\hat{A}|q_2\rangle
=\frac{1}{i\hbar}\langle q_1|\hat{H}\hat{T}\hat{A}|q_2\rangle , \nonumber \\
&=&\frac{1}{i\hbar}H\left( q_1,-i\hbar \frac{\partial}{\partial q_1},t
\right) \langle q_1|\hat{T}\hat{A}|q_2\rangle ,\nonumber  \\
&=&\frac{1}{i\hbar}H\left( q_1,-i\hbar \frac{\partial}{\partial q_1},t
\right) e^{iS_1/\hbar}.
\end{eqnarray}
Eq.\ (16) leads immediately to the desired c-number form of the quantum 
Hamilton-Jacobi equation
\begin{equation}
\left[ H\left( q_1,-i\hbar \frac{\partial}{\partial q_1},t\right) 
+\frac{\partial
S_1(q_1,Q_2,t)}{\partial t}\right] e^{iS_1(q_1,Q_2,t)/\hbar}=0.
\end{equation}
Substitution of $S_2(q_1,P_2,t)$ for $S_1(q_1,Q_2,t)$ generates another 
c-number form of the quantum Hamilton-Jacobi equation. The equations for 
the cases of $S_3(p_1,Q_2,t)$ and $S_4(p_1,P_2,t)$ can be derived through 
a similar process. 

Consider a one-dimensional nonrelativistic quantum system whose 
Hamiltonian is given by
\begin{equation}
H(\hat{q},\hat{p},t)=\frac{\hat{p}^2}{2}+V(\hat{q},t).
\end{equation}
The c-number form of the quantum Hamilton-Jacobi equation (17) for this 
problem becomes
\begin{equation}
\frac{1}{2}\left(\frac{\partial S_1}{\partial q_1}\right)^2
-i\frac{\hbar}{2}\frac{\partial ^2S_1}{\partial q_1^2}+V(q_1,t)
+\frac{\partial S_1}{\partial t}=0.
\end{equation}
We can see clearly that, in the limit $\hbar \rightarrow 0$, the above 
equation reduces to the classical Hamilton-Jacobi equation. The second term 
of Eq.\ (19) represents the quantum effect. We note that it has been known 
from the early days that substitution of 
$\psi (q,t)=e^{iS(q,t)/\hbar}$ into the Schr\"{o}dinger equation gives rise 
to the same Hamilton-Jacobi equation for $S(q,t)$,\footnote{For a stationary
state of a system whose Hamiltonian does not depend explicitly on time, one
may put $S(q,t) = W(q) -Et$ and obtain a differential equation for $W(q)$.
To find a solution to the resulting equation, one may then use the expansion
of $W$ in powers of $\hbar$. This approach has been extensively considered
in connection with the well-known WKB approximation. In the present paper,
the formalism is developed for general nonstationary states (of systems 
that can possibly have time-dependent Hamiltonians).} 
where $S(q,t)$ is interpreted merely as the complex-valued phase of the 
wave function (see, for example, Ref.\cite{schiff}).
The present approach 
more clearly shows the strong analogy between the classical and quantum 
Hamilton-Jacobi theories emphasizing that the quantum Hamilton's principal 
function $S_1$ which is related with the wave function via Eq.\ (14) plays 
the role of the quantum counterpart of the classical generating function. 
Moreover, as discussed later in Sec.\ V, $e^{iS_1/\hbar}$ defined in Eq.\ (14) 
can be interpreted as a propagator under a certain choice of $\hat{A}$.

It may be viewed that the Hamilton-Jacobi equation in the form of Eq.\ (19) 
is no more tractable analytically than the Schr\"{o}dinger equation 
for general potential problems. Nevertheless, it would be possible at least to 
obtain an approximate solution of it using a perturbative method as follows.
Since the solution of Eq.\ (19) is given by the classical Hamilton's 
principal function in the limit $\hbar \rightarrow 0$, we can expand the 
general solution in powers of $\hbar$:
\begin{equation}
S_1=S_1^{(0)}+\hbar S_1^{(1)}+\hbar ^2S_1^{(2)}+\cdots ,
\end{equation}
where $S_1^{(0)}$ is the classical Hamilton's principal function. 
Substituting Eq.\ (20) into Eq.\ (19) and collecting coefficients of the 
same orders in $\hbar$, we can obtain 
\begin{equation}
\frac{1}{2}\left(\frac{\partial S_1^{(0)}}{\partial q_1}\right)^2+V(q_1,t)
+\frac{\partial S_1^{(0)}}{\partial t}=0,
\end{equation}
and
\begin{equation}
\frac{1}{2}\sum_{k=0}^{n}\frac{\partial S_1^{(k)}}{\partial q_1}
\frac{\partial S_1^{(n-k)}}{\partial q_1}
-\frac{i}{2}\frac{\partial ^2S_1^{(n-1)}}{\partial q_1^2}
+\frac{\partial S_1^{(n)}}{\partial t}=0, \hspace{0.5cm} n\geq 1.
\end{equation}
Given the solution $S_1^{(0)}$ of the classical Hamilton-Jacobi equation (21), 
we solve Eq.\ (22) to find $S_1^{(1)}$. $S_1^{(2)}$ can be determined 
subsequently from the knowledge of $S_1^{(0)}$ and $S_1^{(1)}$, and so forth.
We note that Eq.\ (22) is linear in $S_1^{(n)}$ and first-order differential 
in $q_1$ for $S_1^{(n)}$. Thus, from a practical viewpoint, Eqs.\ (21) and
(22) could 
be more advantageous to deal with than Eq.\ (19) as long as
the classical Hamilton's
principal function that is the solution of Eq.\ (21) is readily available. 

The present formalism provides an encouraging point that the well-ordered 
operator counterpart of the quantum Hamilton's principal function gives also 
the solutions of the Heisenberg equations through Eq.\ (2). If we consider
the case $\hat{U}(t)=\hat{T}(t)$, we can obtain in the Heisenberg picture the
relations $(\hat{q}_H,\hat{p}_H)\equiv 
(\hat{T}^\dagger \hat{q}_S\hat{T}, \hat{T}^\dagger \hat{p}_S\hat{T})$
and $(\hat{Q}_H,\hat{P}_H)\equiv (\hat{T}^\dagger \hat{Q}_S\hat{T},
\hat{T}^\dagger \hat{P}_S\hat{T})= (\hat{T}^\dagger \hat{T}\hat{q}_S
\hat{T}^\dagger \hat{T},\hat{T}^\dagger \hat{T}\hat{p}_S\hat{T}^\dagger 
\hat{T})=(\hat{q}_S,\hat{p}_S)$, where we attached the subscript $_S$ and
$_H$ to operators to explicitly denote, respectively, the Schr\"{o}dinger and
the Heisenberg pictures. Thus, when expressed in the Heisenberg 
picture Eq.\ (2) turns into
\begin{equation}
\hat{p}_H=\frac{\partial \bar{S}_1(\hat{q}_H,\hat{q}_S,t)}{\partial \hat{q}_H}, 
\hspace{1cm} 
\hat{p}_S=-\frac{\partial \bar{S}_1(\hat{q}_H,\hat{q}_S,t)}{\partial \hat{q}_S},
\end{equation}
and from these transformation relations we can obtain $\hat{q}_H$ and
$\hat{p}_H$ as functions of time and the initial operators $\hat{q}_S$ and 
$\hat{p}_S$. Obviously, $\hat{q}_H(\hat{q}_S,\hat{p}_S,t)$ and 
$\hat{p}_H(\hat{q}_S,\hat{p}_S,t)$ obtained 
in this way evolve according to the Heisenberg equations. 
\section{Quantum Phase-Space distribution functions and canonical
transformations}

Since our theory of the quantum canonical transformations is formulated with 
the canonical position $\hat{q}$ and momentum $\hat{p}$ variables on an equal
footing, it would be relevant to consider the phase-space picture of quantum 
mechanics, exploiting the distribution functions in relation to the present 
theory.

\subsection{Distribution functions}

For a given density operator $\hat{\rho}$, a general way of defining quantum
distribution functions proposed by Cohen\cite{one5} is that 
\begin{equation}
F^f(q_1,p_1,t)=\frac{1}{2\pi ^2\hbar}\int \int \int dxdydq_2 \langle
q_2+y|\hat{\rho} |q_2-y\rangle f(x,2y/\hbar )e^{ix(q_2-q_1)}
e^{-i2yp_1/\hbar}.
\end{equation}
Various choices of $f(x,2y/\hbar)$ lead to a wide class of quantum 
distribution functions\cite{one6}. To mention only a few, the choice $f=1$ 
produces the well-known Wigner distribution function \cite{one7}, while the 
choice $f(x,2y/\hbar)= e^{-\hbar x^2/4m\alpha -m\alpha y^2/\hbar}$ yields 
the Husimi distribution function that recently has found its application in 
nonlinear dynamical problems\cite{one8}. The transformed distribution 
function is defined in ($Q_1,P_1$) phase space likewise by 
\begin{equation}
G^f(Q_1,P_1,t)=\frac{1}{2\pi ^2\hbar}\int \int \int dXdYdQ_2\hspace{0.7mm}
{_t\langle} Q_2+Y|\hat{\rho} |Q_2-Y\rangle _tf(X,2Y/\hbar )e^{iX(Q_2-Q_1)}
e^{-i2YP_1/\hbar}.
\end{equation}

Our main objective here is to find a relation between the old and the 
transformed distribution functions.  After a straightforward algebra, which 
is displayed in Appendix B, it turns out that the transformation relation 
between the two distribution functions can be expressed as 
\begin{equation}
G^f(Q_1,P_1,t)=\int \int dq_2dp_2\kappa (Q_1,P_1,q_2,p_2,t)
F^f(q_2,p_2,t),
\end{equation}
where the kernel $\kappa$ is given by 
\begin{eqnarray} 
\kappa (Q_1,P_1,q_2,p_2,t)=\frac{1}{2\pi ^3\hbar} \int \int \int
\int \int \int dXdYdQ_2dxdyd\alpha \frac{f(X,2Y/\hbar )}
{f(x,2y/\hbar )} \nonumber \\
\times e^{\frac{i}{\hbar}[F_1(q_2+\alpha -y,Q_2-Y,t)-F_1^*(q_2+\alpha +y,
Q_2+Y,t)]} e^{i[X(Q_2-Q_1)-\alpha x]} e^{\frac{2i}{\hbar}
[yp_2-YP_1]}.
\end{eqnarray}
This expression for the kernel can be further simplified if integrations
in Eq.\ (27) can be performed with a specific choice of the function $f$. 
For instance, the simple choice $f=1$ provides the following kernel for the 
Wigner distribution function, 
\begin{eqnarray}
\kappa (Q_1,P_1,q_2,p_2,t)=\frac{2}{\pi \hbar} \int \int dYdy 
e^{\frac{i}{\hbar} [F_1(q_2-y,Q_1-Y,t)-F_1^*(q_2+y,Q_1+Y,t)]}
e^{\frac{2i}{\hbar}[yp_2-YP_1]}.
\end{eqnarray}
This equation was first derived by Garcia-Calder\'on and Moshinsky\cite{one9}
without employing the idea of the quantum generating function.
Curtright {\it et al.}\cite{one10} also obtained an equivalent expression
in their recent discussion of the time-independent Wigner
distribution functions.

We wish to point out that the quantum canonical transformation described here
is basically different from that considered earlier by Kim and 
Wigner\cite{one1}. 
While the present approach deals with the transformation between operators 
($\hat{q},\hat{p}$) and ($\hat{Q}, \hat{P}$), their approach is about the 
transformation between c-numbers ($q,p$) and ($Q,P$). For the transformation
$Q=Q(q,p,t)$ and $P=P(q,p,t)$, their approach yields for the kernel the 
expression
\begin{equation}
\kappa (Q_1,P_1,q_2,p_2,t)=\delta [Q_1-Q(q_2,p_2,t)]\delta
[P_1-P(q_2,p_2,t)],
\end{equation}
where $Q(q,p,t)$ and $P(q,p,t)$ satisfy the classical Poisson brackets 
relation, $[Q,P]_{PB}=[q,p]_{PB}=1$. The kernels of Eq.\ (28) and Eq.\ (29) 
coincide with each other for the special case of a linear canonical 
transformation, as was shown by Garcia-Calder\'on and Moshinsky\cite{one9}. 
Specifically, for the case of the Wigner distribution function, they showed 
that the linear transformation for operators, $\hat{Q}=a\hat{q}+b\hat{p}$ and 
$\hat{P}=c\hat{q}+d\hat{p}$, and that for c-number variables, $Q=aq+bp$ and 
$P=cq+dp$, yield the same kernel 
$\kappa (Q_1,P_1,q_2,p_2)=\delta [Q_1-(aq_2+bp_2)]\delta [P_1-(cq_2+dp_2)]$.
In general cases, however, Eq.\ (27) and Eq.\ (29) give rise to different 
kernels.  As an example, let us consider the unitary transformation 
$\hat{U}=e^{ig(\hat{q})/\hbar}$ considered in Sec.\ II. The first-type 
quantum generating function has the form 
$e^{iF_1(q_1,Q_2)/\hbar}=e^{ig(q_1)/\hbar}
\delta (q_1-Q_2)$. This nonlinear canonical transformation yields for the 
Wigner distribution function the kernel
\begin{equation}
\kappa (Q_1,P_1,q_2,p_2)= \frac{\delta (Q_1-q_2)}{\pi \hbar} \int 
dy e^{\frac{i}{\hbar}
[g(q_2-y)-g(q_2+y)]} e^{\frac{2i}{\hbar}(p_2-P_1)y}.
\end{equation}
It is apparent that the integral of the above equation cannot generally be 
reduced to the $\delta$-function of Eq.\ (29) except for some trivial cases, 
e.g., $g=$const, $g=q$, and $g=q^2$. Distribution functions other than the 
Wigner distribution function do not usually allow the simple expression for 
the kernel in the form of Eq.\ (29), even if one considers a linear canonical 
transformation.

\subsection{Dynamics}

In this subsection we describe how the quantum Hamilton-Jacobi theory can 
lead to dynamical solutions in the phase-space picture of quantum mechanics. 
For this task, we first consider the time evolution of the transformed 
distribution function in ($Q_1,P_1$) phase space. Differentiating Eq.\ (25) 
with respect to time, we can get
\begin{eqnarray}
\frac{\partial G^f}{\partial t}=\frac{1}{2\pi ^2\hbar}
\int \int \int dXdYdQ_2 \left[ \left( \frac{\partial}
{\partial t}{_t\langle} Q_2+Y| \right) \hat{\rho} |Q_2-Y\rangle _t+
{_t\langle}Q_2+Y|\frac{\partial \hat{\rho}}{\partial t}|Q_2-Y\rangle _t
\right. \nonumber \\
\left. +{_t\langle}Q_2+Y|\hat{\rho} \left( \frac{\partial}{\partial t}
|Q_2-Y\rangle _t\right) \right] f(X,2Y/\hbar)e^{iX(Q_2-Q_1)}
e^{-i2YP_1/\hbar}.
\end{eqnarray}
We now substitute into Eq.\ (31) the time evolution equations
\begin{eqnarray}
\frac{\partial \hat{\rho}}{\partial t}&=&-\frac{i}{\hbar}
[\hat{H},\hat{\rho}], \\
\frac{\partial}{\partial t}{_t\langle}Q_2+Y|&=&{_t\langle}
Q_2+Y|\hat{U}\frac{\partial \hat{U}^\dagger}{\partial t}, \\
\frac{\partial}{\partial t}|Q_2-Y\rangle _t &=&\frac{\partial \hat{U}}
{\partial t}\hat{U}^\dagger |Q_2-Y\rangle _t=-\hat{U}\frac{\partial
\hat{U}^\dagger}{\partial t}|Q_2-Y\rangle _t,
\end{eqnarray} 
where $\hat{H}=H(\hat{q},\hat{p},t)$ is the Hamiltonian governing the 
dynamics of the system, and obtain 
\begin{eqnarray}
\frac{\partial G^f}{\partial t}=\frac{1}{2\pi ^2\hbar}
\int \int \int dXdYdQ_2\hspace{0.7mm} {_t\langle}Q_2+Y|\left( -\frac{i}{\hbar}
[\hat{K},\hat{\rho}]\right) |Q_2-Y\rangle _tf(X,2Y/\hbar)e^{iX(Q_2-Q_1)}
e^{-i2YP_1/\hbar},
\end{eqnarray}
where $\hat{K}=K(\hat{Q},\hat{P},t)$ is just the transformed Hamiltonian 
already defined in Eq.\ (11).  Eq.\ (35) should be compared with the 
following equation that governs the time evolution of the distribution 
function in ($q_1,p_1$) phase space,
\begin{eqnarray}
\frac{\partial F^f}{\partial t}=\frac{1}{2\pi ^2\hbar}
\int \int \int dxdydq_2 \langle q_2+y|\left( -\frac{i}{\hbar}
[\hat{H},\hat{\rho}]\right) |q_2-y\rangle f(x,2y/\hbar)e^{ix(q_2-q_1)}
e^{-i2yp_1/\hbar}.
\end{eqnarray}
We can easily see that, through the quantum canonical transformation, the 
role played by $\hat{H}$ is turned over to $\hat{K}$.

Just as the wave function has a trivial solution in the representation
where the transformed Hamiltonian $K(\hat{Q},\hat{P},t)$ vanishes, so does
the distribution function in the corresponding phase space, as can be seen 
from Eq.\ (35). With the trivial solution $G^f=$\ const., we go back to the 
original space via the inverse of the transformation equation (26) to obtain 
$F^f(q_1,p_1,t)$.  For example, for the case of the Wigner distribution 
function the transformation can be accomplished by
\begin{equation}
F^W(q_1,p_1,t)=\int \int dQ_2dP_2\tilde{\kappa} (q_1,p_1,Q_2,P_2,t)
G^W(Q_2,P_2),
\end{equation}
where $\tilde{\kappa}$ is given in terms of the quantum principal function by
\begin{eqnarray}
\tilde{\kappa} =\frac{2}{\pi \hbar} \int \int dydY e^{\frac{i}{\hbar}
[S_1(q_1+y,Q_2+Y,t)-S_1^*(q_1-y,Q_2-Y,t)]}
e^{\frac{2i}{\hbar}[YP_2-yp_1]}.
\end{eqnarray}
Thus, once the quantum Hamilton-Jacobi equation is solved and the quantum 
principal function $S_1$ is obtained, the dynamics of the distribution 
function, as well as that of the wave function, can be determined.

\section{Boundary conditions and Applications}

Up to this point the whole theory has been developed for the case 
$\hat{U}(t)=\hat{T}(t)\hat{A}$ with $\hat{A}$ taken to be arbitrary 
unless otherwise mentioned. 
To see how the quantum Hamilton-Jacobi theory is used to achieve the
dynamical solutions of quantum problems, it would be 
sufficient, though, to consider the case of $\hat{A}=\hat{I}$, the unit 
operator. This case was considered by Dirac in 
connection with his action principle (see Sec.\ 32 of Ref.\cite{three}). He 
showed that $S_1$ defined by Eq.\ (15) equals the classical action function 
in the limit $\hbar \rightarrow 0$. It should be mentioned that this 
particular case allows the quantum generating functions to attain the 
property that $e^{iS_1/\hbar}$ is the propagator in position space and
$e^{iS_4/\hbar}$ the propagator in momentum space. We will henceforth work 
on the case $\hat{U}(t)=\hat{T}(t)$. The general case 
$\hat{U}(t)=\hat{T}(t)\hat{A}$ will be briefly treated in Appendix C.

Before applying the theory it is necessary to provide some remarks 
concerning the 
quantum Hamilton-Jacobi equation (17) and its solution. First, if 
the Hamiltonian depends only on either $\hat{q}$ or $\hat{p}$, we do not 
need to solve Eq.\ (17). Instead, since the unitary operator has the simple 
form $\hat{U}=\hat{T}=e^{-iH(\hat{q})t/\hbar}$ or $e^{-iH(\hat{p})t/\hbar}$, 
we can obtain $S_1$ directly from Eq.\ (15) by calculating the matrix 
elements of $\hat{U}$. For example, for a free particle, 
$\hat{U}=e^{-i\hat{p}^2t/2\hbar}$, it is convenient to calculate 
$e^{iS_2/\hbar}=\langle q_1|e^{-i\hat{p}^2t/2\hbar}|p_2\rangle$, and we get 
$S_2(q_1,P_2,t)=-\frac{P_2^2t}{2}+q_1P_2+i\frac{\hbar}{2} \ln 2\pi \hbar$.  
Second, in order to solve Eq.\ (17), we need to impose proper boundary 
conditions on $S_1$. Since here we are dealing with unitary transformations, 
we immediately get from the definition of $S_1$ the condition 
\begin{equation}
\int dQ_3 e^{i[S_1(q_1,Q_3,t)-S_1^*(q_2,Q_3,t)]/\hbar} =\delta (q_1-q_2),
\end{equation}
which follows from the calculation of the matrix elements of
$\hat{U}(t)\hat{U}^{\dagger}(t)=\hat{I}$. 
This unitary condition ensures that the well-ordered operator 
counterpart of $S_1$ yields Hermitian operators for $\hat{Q}$ and $\hat{P}$ 
from Eq.\ (2). Mathematically, Eq.\ (17) can have
several solutions, and there is an arbitrariness in the choice of the
new position variable, because any function of the constant of integration  
of Eq.\ (17) can be a candidate for the new position variable. Not all the
possible solutions correspond to the unitary transformations, and from the
possible solutions we choose only those which satisfy Eq.\ (39) and thus give
Hermitian position and momentum operators that are observables. 
These solutions correspond to
the unitary transformations of the type $\hat{U}(t)=\hat{T}(t)\hat{A}$.
Further, from these solutions we single out the one that corresponds to the 
case $\hat{A}=\hat{I}$ by imposing the condition 
$e^{iS_1(q_1,Q_2,t=0)/\hbar}=\delta (q_1-Q_2)$ as an initial 
condition. The appropriate form for $S_2$ corresponding to this condition is
that $e^{iS_2(q_1,P_2,t=0)/\hbar}=\frac{1}{\sqrt{2\pi \hbar}} 
e^{iq_1P_2/\hbar}$. In the limit $\hbar \rightarrow 0$, $S_2$ in this
equation reduces to the correct classical generating function for the
identity transformation, $S_2=q_1P_2$. In solving the Hamilton-Jacobi
equation perturbatively using Eqs.\ (21) and (22), in order to consistently 
satisfy the initial condition, we start with the classical Hamilton's 
principal function $S_1^{(0)}$ that gives at initial
time the relations $q_1=Q_2$ and $p_1=P_2$ from the classical c-number
counterpart of Eq.\ (2). An arbitrary additive constant $c$ to the solution 
of Eq.\ (17) that always appears in the form $S_1+c$ when we deal with a 
partial differential equation such as Eq.\ (19) which contains only partial 
derivatives of $S_1$\cite{one} can also be fixed by the initial condition. 
Depending 
whether boundary conditions can readily be expressed in a simple form, one 
type of the quantum generating function may be favored over another. The
existence and uniqueness of the independent solution of Eq.\ (17) satisfying 
the above conditions can be guaranteed from the consideration of the equation
$e^{iS_1(q_1,Q_2,t)/\hbar}=\langle q_1|\hat{T}(t)|q_2\rangle$, in which
$S_1$ is just given by the matrix elements of $T(t)$. It is clear that
these matrix elements exist and are uniquely defined.

As illustrations of the application of the theory, we consider the following
two simple systems.
 
{\sl Example 1. A particle under a constant force.}

As a first example, let us consider a particle moving under a constant force
of magnitude $a$, for which the Hamiltonian is $\hat{H}=\hat{p}^2/2-a\hat{q}$.
We start with the following classical principal function that is
the solution of Eq. (21),
\begin{equation}
S_1^{(0)}=\frac{(q_1-Q_2)^2}{2t} +\frac{at(q_1+Q_2)}{2}
-\frac{a^2t^3}{24}.
\end{equation}
Substituting $S_1^{(0)}$ into Eq.\ (22) and solving the resulting equation, 
we find that the first order term in $\hbar$ has the general solution
\begin{equation}
S_1^{(1)}=\frac{i}{2}\ln t +f\left( \frac{q_1-Q_2}{t} -\frac{a}{2}t\right) ,
\end{equation}
where $f$ is an arbitrary differentiable function. To satisfy the proper 
boundary condition $e^{iS_1(q_1,Q_2,t=0)/\hbar} =\delta (q_1-Q_2)$,
$f$ and all higher order terms of $S_1$ are chosen to be zero, and the overall 
additive constant to be $c=\hbar \frac{i}{2}\ln i2\pi \hbar$.
By well-ordering terms, we get the generating operator
\begin{equation}
\bar{S_1}(\hat{q},\hat{Q},t)=\frac{\hat{q}^2-2\hat{q}\hat{Q}
+\hat{Q}^2}{2t}+\frac{at}{2}(\hat{q}+\hat{Q})
-\frac{a^2t^3}{24} +\hbar\frac{i}{2}\ln i2\pi \hbar t.
\end{equation}
We can easily check that the operator form of the quantum Hamilton-Jacobi
equation (13) is satisfied by the above generating operator.

From Eq.\ (14) we obtain the wave function
\begin{equation}
\psi ^q(q_1,t)=\int \frac{1}{\sqrt{i2\pi \hbar t}}
e^{\frac{i}{2\hbar t}
[(q_1-Q_2)^2 +at^2(q_1+Q_2)-a^2t^4/12]}\psi ^Q(Q_2)dQ_2.
\end{equation}
Because $\psi ^Q(Q_2)$ is constant in time, we can express it in terms of the
initial wave function. For the present case in which we use the first-type
quantum generating function $S_1$ and $\hat{A}=\hat{I}$, we have simply
$\psi ^q(q_2=Q_2,t=0)=\psi ^Q(Q_2)$. We note that Eq.\ (43) is in exact
agreement with the result of Feynman's path-integral approach\cite{two0}.

For the time evolution of the distribution function, we find from Eq.\ (38)
the following kernel for the Wigner distribution function,
\begin{eqnarray}
\tilde{\kappa} (q_1,p_1,Q_2,P_2,t)&=&\frac{1}{\pi ^2\hbar ^2 t}\int \int
dYdy e^{-\frac{2i}{\hbar}\left(Q_2-q_1+p_1t-\frac{at^2}{2}\right)
\frac{y}{t}} e^{\frac{2i}{\hbar}\left( P_2-\frac{q_1-Q_2-at^2/2}{t}
\right) Y}, \nonumber \\
&=&\delta(Q_2-q_1+p_1t-at^2/2) \delta \left(P_2-\frac{q_1-Q_2-at^2/2}
{t}\right) .
\end{eqnarray}
Substituting Eq.\ (44) into Eq. (37), we obtain
\begin{eqnarray}
F^W(q_1,p_1,t)=F^W(q_1-p_1t+at^2/2,p_1-at,0),
\end{eqnarray}
where use has been made of the relation $F^W(q_1,p_1,t=0)=G^W(q_1,p_1)$.

As has been mentioned, the present Hamilton-Jacobi theory also
provides the solutions of the Heisenberg equations via the transformation
relations between the two sets of canonical operators.
From Eqs.\ (2) and (42) we can obtain
\begin{eqnarray}
\hat{q}_S&=&\hat{Q}_S(t)+\hat{P}_S(t)t+\frac{a}{2}t^2,  \\
\hat{p}_S&=&\hat{P}_S(t) +at.
\end{eqnarray}
In the Heisenberg picture, the above equations become
\begin{eqnarray}
\hat{q}_H(t)&=&\hat{q}_S+\hat{p}_St+\frac{a}{2}t^2,  \\
\hat{p}_H(t)&=&\hat{p}_S +at,
\end{eqnarray}
which are the solutions of the Heisenberg equations.

By setting $a=0$, we can obtain the free particle solution.

{\sl Example 2. The harmonic oscillator}

For the harmonic oscillator whose Hamiltonian is given 
by $\hat{H}=\hat{p}^2/2+\hat{q}^2/2$, the classical Hamilton-Jacobi 
equation (21) can be solved to give the classical principal function
\begin{equation}
S_1^{(0)}=\frac{1}{2}(q_1^2+Q_2^2)\cot t-q_1Q_2\csc t. 
\end{equation}
With the boundary condition $e^{iS_1(q_1,Q_2,t=0)/\hbar}=\delta (q_1-Q_2)$, 
Eq.\ (22) can be solved to give
\begin{equation}
S_1^{(1)}=\frac{i}{2}\ln \sin t,
\end{equation}
and $S_1^{(2)}=\cdots =0$. The additive constant has the form 
$c=\hbar \frac{i}{2}\ln i2\pi \hbar$. The well-ordered generating operator 
is then written as
\begin{equation}
\bar{S_1}(\hat{q},\hat{Q},t)=\frac{1}{2}(\hat{q}^2+\hat{Q}^2)\cot t
-\hat{q}\hat{Q}\csc t +\hbar \frac{i}{2} \ln i2\pi \hbar \sin t.
\end{equation}

The wave function takes the form
\begin{equation}
\psi ^q(q_1,t)=\int \frac{1}{\sqrt{i2\pi \hbar \sin t}}
e^{\frac{i}{2\hbar \sin t}
[(q_1^2+Q_2^2)\cos t -2q_1Q_2]}\psi ^q(Q_2,0)dQ_2,
\end{equation}
and the kernel and the distribution function are given respectively by
\begin{equation}
\tilde{\kappa} (q_1,p_1,Q_2,P_2,t)=\delta (Q_2-q_1\cos t +p_1\sin t )
\delta (P_2+Q_2\cos t -q_1\csc t),
\end{equation}
and
\begin{equation}
F^W(q_1,p_1,t)=F^W(q_1\cos t-p_1\sin t,q_1\sin t+p_1\cos t,0).
\end{equation}
This equation shows that the Wigner distribution function for the harmonic
oscillator rotates clockwise in phase space.

The quantum Hamilton-Jacobi equation for other types of generating operators
can be solved by a similar technique. For instance, we can obtain the 
following solution for the second-type generating operator,
\begin{equation}
\bar{S_2}(\hat{q},\hat{P},t)=-\frac{1}{2}(\hat{q}^2+\hat{P}^2)\tan t
+\hat{q}\hat{P}\sec t+\hbar \frac{i}{2} \ln 2\pi \hbar \cos t.
\end{equation}

The solutions of the Heisenberg equations can be obtained from Eqs.\ (2) and 
(52) (or Eqs. (3) and (56)). In the Heisenberg picture we have 
\begin{eqnarray}
\hat{q}_H(t)&=&\hat{q}_S\cos t+\hat{p}_S\sin t, \\
\hat{p}_H(t)&=&-\hat{q}_S\sin t+\hat{p}_S\cos t.
\end{eqnarray}

It should be mentioned that, even though we restricted our discussion
in this section only to the case $\hat{A}=\hat{I}$ by imposing
the special initial condition, it is very probable that another choice of 
$\hat{A}$ satisfying the quantum Hamilton-Jacobi equation happens to be more
readily obtainable. In that case, the initial condition that
is derived from $e^{iS_1(q_1,Q_2,0)/\hbar}=\langle q_1|\hat{A}|q_2\rangle$ is 
of course different from that described above. As an example, for the
harmonic oscillator, we presented a different solution for $S_1$ in Appendix C 
where the unitary operator $\hat{A}$ corresponds to the transformation that 
interchanges the position and momentum operators.

\section{Concluding remarks}

We wish to give some final remarks concerning the quantum Hamilton-Jacobi 
theory. In this approach, the quantum Hamilton-Jacobi equation takes the 
place of the time-dependent Schr\"{o}dinger equation for solving dynamical 
problems, and the quantum Hamilton's principal function $S_1$ that is the 
solution of the former equation gives the solution of the latter equation 
through Eq.\ (14).  As mentioned in Sec.\ V, $e^{iS_1/\hbar}$ becomes the 
propagator in position space for the case $\hat{A}=\hat{I}$. To find the 
propagator, Feynman's path-integral approach divides the time difference 
between a given initial state and a final state into infinitesimal time 
intervals, and then lets the quantum generating function for the 
infinitesimal transformation equal the classical action function plus
a proper additive constant that vanishes in the limit $\hbar \rightarrow
0$, and finally takes the sum of the infinitesimal 
transformations. On the other hand, the present approach seeks the 
quantum generating function that 
directly transforms the initial state to the final state.
The present formalism gives also the solutions of the Heisenberg equations 
through the transformation relations which in the Heisenberg picture can
be expressed as Eq.\ (23). 
In conclusion, it is clear that the present approach, which has its origin 
in Dirac's canonical transformation theory, helps better comprehend the 
interrelations among the existing different formulations of quantum mechanics.

Finally, one more remark may be worth making as to the extent to which the
quantum Hamilton-Jacobi theory can stretch the range of its validity. 
Even though our work here deals with the
unitary transformation to ensure that the new operators become Hermitian,
and hence observables, the main idea presented in this paper could be
extended so as to include the non-unitary transformation that deals with
non-Hermitian operators. The theory would then have the form of the
quantum Hamilton-Jacobi equation, but it would be associated with  
different types of 
transformations, such as $\hat{U}(t)=\hat{T}(t)\hat{B}$ where $\hat{U}(t)$ and 
$\hat{B}$ are not unitary. However, it may then be necessary to pay particular 
attention 
and care to the completeness of the eigenstates of the new operators 
$\hat{Q}$ and $\hat{P}$, for the property is crucial to several relations 
derived and has been used implicitly throughout the paper.

\appendix

\section{Proof of the equivalence of equations (11) and (12)} 
Eq.\ (12) can be derived from Eq.\ (11) by considering the matrix element 
of the second term on the right hand side of Eq.\ (11) as follows,
\begin{eqnarray}
\langle q_1|i\hbar \hat{U}\frac{\partial \hat{U}^\dagger}{\partial t}| q_2 
\rangle &=& -\langle q_1|i\hbar \frac{\partial \hat{U}}{\partial t}
\hat{U}^\dagger |q_2 \rangle , \\
&=&-\int dq_3\langle q_1|i\hbar \frac{\partial \hat{U}}{\partial t} 
|q_3\rangle \langle q_3|\hat{U}^\dagger |q_2 \rangle , \\
&=&\int dQ_3\left(-i\hbar \frac{\partial}{\partial t}\langle
q_1|\hat{U}|q_3\rangle \right) {_t\langle}Q_3|q_2\rangle ,
\end{eqnarray}
where the identity $\hat{U}\hat{U}^\dagger =\hat{I}$ is used to obtain 
Eq.\ (A1).
Using the definition of the quantum generating function (1), we can obtain
\begin{eqnarray}
\langle q_1|i\hbar \hat{U}\frac{\partial \hat{U}^\dagger}{\partial t}| q_2 
\rangle &=& \int dQ_3 \frac{\partial F_1(q_1,Q_3)}{\partial t}
\langle q_1|Q_3\rangle _t\hspace{0.7mm} {_t\langle}Q_3|q_2\rangle , \\
&=& \int dQ_3\langle q_1|\frac{\partial \bar{F}_1(\hat{q}, \hat{Q},t)}
{\partial t} |Q_3\rangle _t\hspace{0.7mm} {_t\langle} Q_3|q_2\rangle , \\
&=&\langle q_1|\frac{\partial \bar{F}_1(\hat{q},\hat{Q},t)}
{\partial t}|q_2 \rangle .
\end{eqnarray}
Since $i\hbar \hat{U}\frac{\partial \hat{U}^\dagger}{\partial t}$ and 
$\frac{\partial \bar{F}_1(\hat{q},\hat{Q},t)}{\partial t}$ have the 
same matrix element, we conclude that the two operators are identical.  

\section{Derivation of the kernel $\kappa$}
In order to find the relation between $F^f$ and $G^f$, we first make use of
the completeness of the eigenvectors of $\hat{q}$ in Eq.\ (25) and get 
\begin{eqnarray}
G^f(Q_1,P_1,t)&=&\frac{1}{2\pi ^2\hbar}\int \int \int \int \int 
dXdYdQ_2dq_3dq_4\hspace{0.7mm} {_t\langle}Q_2+Y|q_3\rangle \langle q_3|
\hat{\rho} |q_4\rangle \langle q_4|Q_2-Y\rangle _t  \nonumber \\
&&\times f(X,2Y/\hbar )e^{iX(Q_2-Q_1)} e^{-i2YP_1/\hbar}.
\end{eqnarray}
Changing variables with $q_5=(q_3+q_4)/2$ and $y=(q_3-q_4)/2$, and using the 
relation $\int dy'dq_6\delta (y-y')\delta (q_5-q_6) g(y',q_6)=g(y,q_5)$ where
the $\delta$-functions are written in the forms
\begin{equation}
\delta (y-y')=\frac{1}{\pi \hbar}\int dp_3e^{-2ip_3(y-y')/\hbar},
\end{equation}
and
\begin{equation}
\delta (q_5-q_6)=\frac{1}{2\pi}\int dxe^{ix(q_6-q_5)},
\end{equation}
we obtain
\begin{eqnarray}
G^f(Q_1,P_1,t)&=&\frac{1}{2\pi ^4\hbar ^2}\int \int \int \int 
\int \int \int \int \int dXdYdQ_2dq_5dydy'dp_3dq_6dx 
f(X,2Y/\hbar ){_t\langle}Q_2+Y|q_5+y'\rangle  \nonumber \\
&& \times \langle q_6+y|\hat{\rho} |q_6-y\rangle \langle 
q_5-y'|Q_2-Y\rangle _t e^{iX(Q_2-Q_1)} 
e^{-i2YP_1/\hbar} e^{-2ip_3(y-y')/\hbar} e^{ix(q_6-q_5)}.
\end{eqnarray}
Next, we multiply this equation by
\begin{equation}
\int \int dx''dy'' \frac{f(x,2y/\hbar )}{f(x'',2y''/\hbar )}\delta 
(x''-x)\delta (y''-y)=1,
\end{equation}
where 
\begin{eqnarray}
\delta (x''-x)\delta (y''-y)=\frac{1}{2\pi ^2\hbar} \int \int
d\alpha d \beta e^{-i\alpha (x''-x)} e^{-2i\beta (y''-y)/ \hbar}.
\end{eqnarray}
In the resulting equation, we replace the integrations over $q_5$ and $p_3$,
respectively, with those over $q_2=q_5-\alpha$ and $p_2=p_3-\beta$, and
then integrate over $\beta$ and $y'$. We then obtain
\begin{eqnarray}
G^f(Q_1,P_1,t)&=&\int \int dq_2dp_2\bigg[ \frac{1}{2\pi ^3\hbar}
\int \int \int \int \int \int dXdYdQ_2dx''dy''d\alpha
{_t\langle} Q_2+Y| q_2+\alpha +y''\rangle  \nonumber \\
&& \times \langle q_2+\alpha -y''|Q_2-Y \rangle _t 
\frac{f(X,2Y/\hbar )}{f(x'',2y''/\hbar )}
e^{iX(Q_2-Q_1)} e^{-2iYP_1/\hbar} e^{-i\alpha x''} e^{2ip_2y''/\hbar} \bigg] 
\nonumber \\
&& \times \left[ \frac{1}{2\pi ^2\hbar} \int \int \int dxdydq_6
\langle q_6+y|\hat{\rho} |q_6-y\rangle f(x,2y/\hbar )
e^{ix(q_6-q_2)}e^{-2iyp_2/\hbar} \right] ,
\end{eqnarray}
which leads immediately to Eq.\ (26). 
 
\section{Solutions of the quantum Hamilton-Jacobi equation}
In Sec.\ V we considered the case $\hat{A}=\hat{I}$ only for convenience, 
because,
as can be noticed from the two examples of Sec.\ V, this case not only gives
a simple relation between the initial wave function $\psi ^q(q_1,0)$
(distribution function $F^f(q_1,p_1,0)$) in $q$-representation and the constant
wave function $\psi ^Q(Q_2)$ (distribution function $G^f(Q_2,P_2)$) in 
$Q$-representation, but also makes it easy to express the new 
operators $(\hat{Q}_H,\hat{P}_H)$ in the Heisenberg picture in terms of the 
old operators $(\hat{q}_S,\hat{p}_S)$ in the Schr\"{o}dinger picture.
In doing so, we required the solution to satisfy the
specific initial condition described in the first part of Sec.\ V. This
specialization is of course not of necessity, and if this initial condition
is discarded with the unitary condition of the transformation still retained,
we have a group of general solutions each member of which corresponds to a 
specific choice of $\hat{A}$. As an illustration, we present below 
another possible 
solution of the quantum Hamilton-Jacobi equation (19) for the harmonic 
oscillator that belongs to unitary transformations of the type 
$\hat{U}(t)=\hat{T}(t)\hat{A}$, 
\begin{equation}
\bar{S_1}(\hat{q},\hat{Q},t)=-\frac{1}{2}(\hat{q}^2+\hat{Q}^2)\tan t
+\hat{q}\hat{Q}\sec t+\hbar \frac{i}{2} \ln 2\pi \hbar \cos t.
\end{equation}
Equation (C1) satisfies at initial time $e^{iS_1(q_1,Q_2,0)/\hbar}=\frac{1}
{\sqrt{2\pi \hbar}}
e^{iq_1Q_2/\hbar}$, which equals $\langle q_1|\hat{A}|q_2\rangle$, i.e.,  
the matrix element of the unitary operator $\hat{A}$. 
In this case, the operator
$\hat{A}$ corresponds to the exchange transformation which generates
the transformation relations 
\begin{eqnarray}
\hat{p}&=&\frac{\partial \bar{S_1}(\hat{q},\hat{Q},0)}{\partial \hat{q}} 
=\hat{Q}, \\
\hat{P}&=&-\frac{\partial \bar{S_1}(\hat{q},\hat{Q},0)}{\partial \hat{Q}} 
=-\hat{q}.
\end{eqnarray}

\end{document}